\documentstyle[12pt,cite]{article}

\def \beq {\begin{equation}}
\def \eeq {\end{equation}}
\def \endp{\rule{1ex}{1ex}}

\begin{document}

\begin{titlepage}

\begin{flushright}
{\bf IFUSP/P-1068}
\end{flushright}

\begin{center}
{\Large{\bf Volume-forms and Minimal Action Principles in Affine Manifolds }}

\vspace{1cm}

{\large{\bf Alberto Saa}}

\vspace{.5cm}

{\it Instituto de F\'{\i}sica \\
Universidade de S\~ao Paulo, Caixa Postal 20516 \\
01498-970 S\~ao Paulo,  Brazil\\}

\vspace{1cm}

{\bf Abstract}
\end{center}
{\small
Through the analyses of volume-forms in differentiable manifolds, it is shown
that the usual way of defining minimal action principles for field theory
on curved space-times
 is not appropriate on non-riemannian manifolds. An alternative approach,
based in a new volume-form, is proposed and
confronted with the standard one.
 The new volume element is explicitly used in
the study of Einstein-Cartan theory of gravity and its relation to
string theory,
 in connection with some recent results on the subject.
}

\vspace{2cm}

\begin{flushleft}
\end{flushleft}

\end{titlepage}

\newpage

\section{Introduction}

This work discusses the problem  of volume definition in differentiable
manifolds and its relation with minimal action principles.
Action principles are the starting point for several models in Physics, and
they usually are formulated in non-euclidean (non-minkowskian) manifolds.
In all these models,  lagrangians  are integrated
over the manifold, and under the variational principle hypothesis the dynamical
equations are gotten. Whenever such lagrangians are scalars, they are
integrated by using a covariant volume elements so that in a given coordinate
system one has
\beq
\label{act}
S = \int {\cal L}d{\rm vol}=\int {\cal L} \sqrt{|g|} d^nx.
\eeq
Usually,
the quantity $\sqrt{|g|}$ is introduced  with the argument that it
makes the euclidean volume element $d^nx$ covariant. Typically, the cases where
gravitation is incorporated into field theory are described by actions of the
type (\ref{act}). In these cases, space-time can be assumed to be
non-riemannian manifolds, as for example in the Einstein-Cartan theory of
gravity\cite{hehl}.

It will be shown that, for affine manifolds, there are natural compatibility
conditions that a volume element should obey. These conditions will
be used in order
to construct compatible volume elements for
general affine manifolds, and as one of their consequences,
one will have that for
non-riemannian manifolds the usual volume element (\ref{act}) is not the
more appropriate one. This is a very important point, since (\ref{act})
is  adopted for non-riemannian manifolds as well.

Compatible volume elements demand some restrictions on the non-rie\-man\-ni\-an
structure of the manifold, or in other words, general affine
manifolds may not
admit compatible volume elements.
In these cases, it is not clear how to choose a volume element and
consequently, how to define  action principles for field theory.

In this work, ${\cal M}$ is an $n$-dimensional $C^\infty$ differentiable
oriented manifold, and $\Omega^m({\cal M})$ the space of differential
$m$-forms on it. ${\cal M}$ is called an affine manifold if it is endowed
with a linear connection $\Gamma^\alpha_{\beta\gamma}$, which is used to
define the covariant derivative of  tensor valued differential
forms
\beq
\label{tform}
D\Pi^\alpha_\beta = d\Pi^\alpha_\beta +
\omega^\alpha_\rho\wedge\Pi^\rho_\beta -
\omega^\rho_\beta\wedge\Pi_\rho^\alpha - w\omega\wedge\Pi^\alpha_\beta,
\eeq
where $\omega^\alpha_\beta=\Gamma^\alpha_{\mu\beta}dx^\mu$,
      $\omega = \Gamma^\alpha_{\alpha\mu}dx^\mu$, and $w$ is the weight
of $\Pi^\alpha_\beta$.
Of course, (\ref{tform}) can be extended for tensor valued forms of any
rank. Considering a ($p,q$) relative tensor of
weight $w$, $A^{(\alpha_1..\alpha_p)}_{(\beta_1..\beta_p)}$,
as $p\times q$ differential $0$-forms
 one
gets from (\ref{tform})
the usual formula for the covariant derivative of relative tensors:
\begin{eqnarray}
&&DA^{(\alpha)}_{(\beta)} = \left(D_\mu A^{(\alpha)}_{(\beta)}\right)dx^\mu =
\\&&= \left(\partial_\mu  A^{(\alpha)}_{(\beta)} +
\sum_{i=1}^p\Gamma^{\alpha_i}_{\mu\rho}
A^{(\alpha_1..\rho..\alpha_p)}_{(\beta )}
-\sum_{j=1}^q\Gamma^\rho_{\mu\beta_j}A^{(\alpha)}_{(\beta_1..\rho..\beta_q)}
-  w\Gamma^\rho_{\rho\mu}A^{(\alpha)}_{(\beta)}\right)dx^\mu. \nonumber
\end{eqnarray}
We will assume also that a metric tensor $g_{\alpha\beta}$ is defined on
${\cal M}$ so that
\beq
ds^2 = g_{\alpha\beta}dx^\alpha dx^\beta.
\label{metric}
\eeq

The anti-symmetrical part of the affine connection
\begin{equation}
S_{\alpha\beta}^{\ \ \gamma} = \frac{1}{2}
\left(\Gamma_{\alpha\beta}^\gamma-\Gamma_{\beta\alpha}^\gamma \right),
\label{torsion}
\end{equation}
defines a new tensor, the torsion tensor $S_{\alpha\beta}^{\ \ \gamma}$. It
is well known that the $n^3$ independent components of the affine connection
can be written as function of the metric (\ref{metric}),
the torsion (\ref{torsion}), and the non-metricity
tensor, defined by
\beq
N_{\alpha\beta\gamma} = D_\alpha g_{\beta\gamma}.
\eeq
The expression for the connection as a function of such quantities is
\beq
\Gamma^{\alpha}_{\beta\gamma} = \left\{^\alpha_{\beta\gamma}\right\} -
K_{\beta\gamma}^{\ \ \alpha} + V^\alpha_{\ \beta\gamma},
\eeq
where $\left\{^\alpha_{\beta\gamma}\right\}$ are the Christoffel symbols,
$K_{\alpha\beta}^{\ \ \gamma}$ is the contorsion tensor,
\beq
K_{\alpha\beta}^{\ \ \gamma} =  -S_{\alpha\beta}^{\ \ \gamma}
+ S_{\beta\ \alpha}^{\ \gamma\ } - S_{\ \alpha\beta}^{\gamma\ \ },
\eeq
and $V_{\alpha\beta\gamma}$ is given by:
\beq
V_{\alpha\beta\gamma} = \frac{1}{2}\left(
N_{\alpha\beta\gamma} - N_{\gamma\alpha\beta} - N_{\beta\gamma\alpha}
\right).
\eeq
For simplicity, the traces of $S^{\ \ \rho}_{\nu\mu}$ and
$V^{\ \ \rho}_{\nu\mu}$  will be denoted by
$S_\mu$ and $V_\mu$ respectively,
\begin{eqnarray}
S_\mu &=& S^{\ \ \rho}_{\rho\mu}, \nonumber \\
V_\mu &=& V^{\ \ \rho}_{\rho\mu}.
\end{eqnarray}

An affine manifold
${\cal M}$ is called a Riemann--Cartan manifold if $N_{\alpha\beta\gamma}=0$,
and a riemannian one if $N_{\alpha\beta\gamma}=0$ and
$S_{\alpha\beta\gamma}=0$. In all these cases, the connection is said to be
metric-compatible.

\section{Volume-forms in Affine Manifolds}
\label{sec2}

In this section, it will be introduced the notion of compatible
volume-form,
and it will be derived the necessary conditions that an affine manifold must
obey in order to be possible the definition of a
compatible volume element on it.

\noindent{\bf Definition 1. }{\em A volume-form on ${\cal M}$ is a nowhere
vanishing $n$-form $v \in \Omega^n({\cal M})$.}\cite{nakahara}

A volume-form, in general, can be constructed by using $n$ linearly independent
$1$-forms ($\theta^1\wedge ...\wedge\theta^n\neq 0$) and
non-vanishing $0$-forms,
\beq
v=f\theta^1\wedge ...\wedge\theta^n,
\label{def1}
\eeq
We will assume that $f$ is a positive non-vanishing $C^\infty$ scalar
function. The volume-form (\ref{def1})
defines a volume element on ${\cal M}$.
If $\{\theta^i\}$ is assumed to be
an orthonormal set of 1-forms, one has the following
expression for the volume-form in general coordinates $\{x^i\}$
\beq
\label{defa}
v=f(x)\sqrt{|g|}dx^1\wedge...\wedge dx^n,
\eeq
as one can check using that if $\Lambda$ is the change of basis matrix
($\theta^i=\Lambda^i_jdx^j$),  then
$v=f(x)\det(\Lambda)dx^1\wedge...\wedge dx^n$. The determinant $\det(\Lambda)$
can be obtained by using the transformation properties of the metric tensor,
and that $\{\theta^i\}$ is orthonormal.

Using (\ref{defa}) we have for the total volume of the manifold
\beq
\int v = \int f(x)\sqrt{|g|}dx^1dx^2...dx^n = \int d{\rm vol}.
\eeq
The usual volume element for riemannian space-times is obtained by choosing
$f(x)=1$.

In an affine manifold, one can require certain compatibility conditions
between the affine connection and the volume-form. For a differentiable
manifold  with volume-form $v$, one usually defines the divergence
of a vector field $A$, ${\rm div\ }A$, by \cite{kob}
\beq
({\rm div\ }A)v = \pounds_{\!{}_A}v,
\label{div}
\eeq
where $\pounds_{\!{}_A}$ is the Lie derivative along
 the direction $A$. However, if
the manifold is endowed with an affine connection, we can define the
divergence of a vector field in a very natural way by using the covariant
derivative,
\beq
{\rm div}_{\!{}_\Gamma} A = D_\mu A^\mu.
\label{divg}
\eeq
One can use (\ref{div}) and (\ref{divg}) to define a criterion of compatibility
between the affine connection and the volume-form.

\noindent{\bf Definition 2. }{\em A volume-form $v$ is compatible with the
affine connection if
\beq
\label{def2}
 \pounds_{\!{}_A} v = (D_\mu A^\mu)v ,
\eeq
for any vector field $A$.}

\noindent One can check that the riemaniann volume-form
$\omega=\sqrt{|g|}dx^1\wedge...\wedge dx^n$ and the Christoffel symbols are
compatible.
It can be inferred also that the volume-form
$\omega$
is not compatible with the connection for a Riemann--Cartan manifold,
\beq
\label{rc1}
 \pounds_{\!{}_A} \omega = \left(
D_\mu A^\mu - 2S_\mu A^\mu
\right)\omega\ne (D_\mu A^\mu)\omega.
\eeq

The incompatibility of the usual volume-form and the connection for
non-riemannian manifolds introduces a new question: Is it possible
to define a volume-form compatible with the connection for non-riemannian
manifolds? The answer is that sometimes it is, as we will see.

\noindent{\bf Proposition 1. }{\em An affine manifold admits a volume-form
compatible with the connection only if the form $(V_\beta + 2S_\beta)dx^\beta$
is exact.}

\noindent{\bf Proof: } In an affine manifold with volume-form $v$ one has
\beq
\pounds_{\!{}_A} v = \left[
A^\mu D_\mu\left(f(x)\sqrt{|g|}\right)
+ f(x)\sqrt{|g|}D_\mu A^\mu\right]dx^1\wedge...\wedge dx^n,
\eeq
and in order to get (\ref{def2}) for arbitrary $A^\mu$ one needs
\begin{eqnarray}
D_\mu(f(x)\sqrt{|g|}) &=& \sqrt{|g|}\partial_\mu f(x)
+ f(x)\partial_\mu\sqrt{|g|}
 -\left\{^\rho_{\rho\mu} \right\}f(x)\sqrt{|g|} \nonumber \\ &-&
(V_{\mu}+2S_{\mu})f(x)\sqrt{|g|} = 0,
\end{eqnarray}
which leads to
$\partial_\mu\ln f(x) = V_{\mu}+2S_{\mu}$, or equivalently
\beq
\label{poin}
V_{\mu}dx^\mu+2S_{\mu}dx^\mu = d\ln f(x) ,
\eeq
where $d$ stands to exterior derivative.\endp

{}From (\ref{poin})
 the $1$-form in question is closed as consequence of Poincare's lemma.
If the form $(V_\beta + 2S_\beta)dx^\beta$ is not closed, the affine manifold
does not admit a compatible volume-form. A connection compatible
volume-form in an affine manifold will be given by
\beq
v = e^{2\Theta}\sqrt{|g|}dx^1\wedge ... \wedge dx^n,
\eeq
where $\partial_\mu \Theta = S_\mu + \frac{1}{2}V_\mu$, and we also have
that
\beq
\label{contconn}
\Gamma^\rho_{\rho\mu} = \partial_\mu \ln \left(e^{2\Theta}\sqrt{|g|}\right).
\eeq

\noindent{\bf Proposition 2. }{\em If an affine manifold is endowed with a
volume-form
compatible with the connection, one has the following generalized Gauss
formula
\beq
\label{gauss}
\int_{\cal M} D_\mu A^\mu d\mu = \int_{\partial\cal M}A^\mu d\Sigma_\mu,
\eeq
where $d\Sigma_\mu$ is the compatible surface element, given by:}
\beq
d\Sigma_\mu = \frac{e^{2\Theta}\sqrt{|g|}}{(n-1)!}
\varepsilon_{\mu\alpha_2\alpha_3...\alpha_n}
dx^{\alpha_2}\wedge dx^{\alpha_3}...\wedge dx^{\alpha_n}.
\eeq

\noindent{\bf Proof: } By choosing a form
$\omega = \frac{e^{2\Theta\sqrt{|g|}}}{(n-1)!}\varepsilon_{\alpha_1\alpha_2...
\alpha_n}A^{\alpha_1}dx^{\alpha_2}\wedge...\wedge dx^{\alpha_n}$, using
(\ref{contconn}) and Stokes' theorem
\beq
\int_{\cal M} d\omega = \int_{\partial\cal M}\omega,
\eeq
one gets (\ref{gauss}).\endp

For  riemannian manifolds, the connection compatible volume-form can be
obtained by using the Hodge (${}^*$) operator.
 The (${}^*$)  is a linear operator\cite{nakahara}
\begin{equation}
{}^* : \Omega^m({\cal M}) \rightarrow  \Omega^{n-m}({\cal M}),
\end{equation}
which for a Riemannian manifold has the following action on a basis vector
of $\Omega^m({\cal M})$
\begin{equation}
{}^* ( dx^{\alpha_1} \wedge dx^{\alpha_2} \wedge ... \wedge dx^{\alpha_m}) =
\frac{\sqrt{|g|}}{(n-m)!}
\varepsilon^{\alpha_1 ... \alpha_m}_
{\ \ \ \ \ \ \ \beta_{m+1}...\beta_n}
dx^{\beta_{m+1}}\wedge ... \wedge dx^{\beta_n},
\label{hodge}
\end{equation}
where $\varepsilon_{\alpha_1 ... \alpha_n}$ is the totally anti-symmetrical
symbol, and
$\varepsilon^{\alpha_1 ... \alpha_m}_{\ \ \ \ \ \ \ \beta_{m+1}...\beta_n}$
is constructed by using the metric tensor.
The action of (\ref{hodge}) on the basis vector of
$\Omega^0({\cal M})$ gives
\begin{equation}
{}^* 1 = \frac{\sqrt{|g|}}{n!} \varepsilon_{\alpha_1 ... \alpha_n}
dx^{\alpha_1}\wedge ... \wedge dx^{\alpha_n} = \sqrt{|g|} d^n x,
\label{ele}
\end{equation}
which is the compatible volume element for a Riemannian manifold.

\noindent{\bf Proposition 3. }{\em If an affine manifold ${\cal M}$
admits a connection
compatible volume-form, it can be obtained  using the modified Hodge
$({}^*)$ operator given by}
\beq
\label{mhodge}
{}^* ( dx^{\alpha_1} \wedge dx^{\alpha_2} \wedge ... \wedge dx^{\alpha_m}) =
\frac{e^{2\Theta}\sqrt{|g|}}{(n-m)!}
\varepsilon^{\alpha_1 ... \alpha_m}_
{\ \ \ \ \ \ \ \beta_{m+1}...\beta_n}
dx^{\beta_{m+1}}\wedge ... \wedge dx^{\beta_n}.
\eeq

\noindent{\bf Proof: } It is clear that the action of (\ref{mhodge}) in the
basis vector for $\Omega^0({\cal M})$ gives:
\beq
{}^* 1 = \frac{e^{2\Theta}\sqrt{|g|}}{n!} \varepsilon_{\alpha_1 ... \alpha_n}
dx^{\alpha_1}\wedge ... \wedge dx^{\alpha_n} = e^{2\Theta}\sqrt{|g|} d^n x,
\eeq
which is the compatible volume-form.\endp

\section{Physical Applications}

In this section,
the results of Sect. \ref{sec2} will be used in the study
of Einstein-Cartan theory of
gravity and its relation with string theory.

In the Einstein-Cartan theory of gravity, the space-time is assumed to be a
Riemann-Cartan manifold.
It is in accordance with the experimental data
and it has also theoretical importance, since it is the theory
that arises from the local gauge theory for the Poincar\'e's
group \cite{hehl}.
The dynamical equations for such theory are gotten via
a minimal action principle of the type (\ref{act}), and the lagrangians
for external fields are usually
obtained from the minkowskian ones by  minimal coupling
procedure.

We propose that, instead of (\ref{act}), the action formulation for
Einstein-Cartan theory shall use the compatible volume element of
section \ref{sec2}. Of course  the restriction that the
space-time
manifold admits such volume element (Proposition 1)
 is implicit. The dynamical
equations will be different from the usual ones, and probably a final
answer on which is more appropriate can be given only by experimental facts.
However, there are theoretical evidences on behalf of the new action, and they
will be pointed out in this section.

According to our hypothesis, the Einstein-Cartan gravity equations shall be
obtained from a Hilbert-Einstein action using the compatible volume element,
\begin{eqnarray}
\label{action}
S &=& -\int e^{2\Theta}\sqrt{-g} d^4x {\cal R} \\
  &=& -\int e^{2\Theta}\sqrt{-g} d^4x \left(R + 4\partial_\mu\Theta
\partial^\mu\Theta - K_{\nu\rho\alpha}K^{\alpha\nu\rho} \right)
+{\rm surf.\ terms},\nonumber
\end{eqnarray}
where the generalized Gauss' formula (\ref{gauss}) was used. In (\ref{action}),
${\cal R}$ is the scalar of curvature of the Riemann-Cartan manifold,
calculated by the contraction of the curvature tensor obtained using the
full connection, and $R$ is the usual riemannian scalar of curvature,
calculated from the Christoffel symbols.
The following conventions are adopted:
${\rm sign}(g_{\mu\nu})=(+,-,-...)$,
$R_{\alpha\nu\mu}^{\ \ \ \beta} = \partial_\alpha\Gamma_{\nu\mu}^\beta
+ \Gamma_{\alpha\rho}^\beta\Gamma_{\nu\mu}^\rho -
(\alpha\leftrightarrow\nu)$, and
$R_{\nu\mu}=R_{\alpha\nu\mu}^{\ \ \ \alpha}$.

The similarity between (\ref{action}) and the action for the dilaton
gravity[\citen{callan,green}]
is surprising. The ``torsion potential'' $\Theta$ can
be identified with the dilaton field, and (\ref{action}) can provide a
geometrical interpretation for the dilaton gravity\cite{saa3}.
 Another feature of the
proposed action is that, due to the peculiar $\Theta$-dependence
of the action (\ref{action}), the trace of the
torsion tensor can propagate, {\em i.e.} it can exist non-vanishing solutions
for torsion in the vacuum. The torsion mediated interactions are not of
contact type anymore. The traceless part of the torsion tensor will be zero
in the vacuum, as in the usual Einstein-Cartan theory.

Another problem that can be analyzed with the results of Sect. \ref{sec2} is
the description of Maxwell fields on Riemann-Cartan space-times.
In order to study Maxwell's equations in a metric differentiable manifold,
we introduce a fundamental quantity, the (local)
electromagnetic potential $1$-form
\begin{equation}
A=A_\alpha dx^\alpha,
\label{potvet}
\end{equation}
and from the potential $1$-form we can define
the Faraday's $2$-form
\begin{equation}
F = dA = \frac{1}{2} F_{\alpha\beta}\, dx^\alpha\wedge dx^\beta,
\label{faraday}
\end{equation}
where $F_{\alpha\beta}=\partial_\alpha A_\beta- \partial_\beta A_\alpha$ is
the
usual electromagnetic tensor.

The homogenous Maxwell's equations arise naturally due to the definition
(\ref{faraday}) as a consequence of Poincar\'e's lemma
\begin{equation}
dF = d(dA) = \frac{1}{2}\partial_\gamma F_{\alpha\beta}\,
dx^\gamma\wedge dx^\alpha \wedge dx^\beta = 0,
\label{1st}
\end{equation}
and in terms of components we have
\begin{equation}
\partial_{[\gamma} F_{\alpha\beta]} = 0,
\label{1stc}
\end{equation}
where ${}_{[\ \ \ ]}$ means antisymmetrization.

The non-homogenous equations in Minkowski space-time are given by
\begin{equation}
d {}^*\!F = 4\pi {}^*\! J,
\label{2nd}
\end{equation}
where $J = J_\alpha dx^\alpha$ is the current 1-form, and (${^*}$) is the
Hodge operator in Minkowski space-time. ${}^*\!J$ and ${}^*\!F$ are given by:
\begin{eqnarray}
{}^*\!J &=& \frac{1}{3!} \varepsilon_{\alpha\beta\gamma\delta}J^\alpha
dx^\beta\wedge dx^\gamma \wedge dx^\delta, \\
{}^*\!F &=& \frac{1}{4}\varepsilon_{\alpha\beta\gamma\delta} F^{\gamma\delta}\,
dx^\alpha \wedge dx^\beta.
\end{eqnarray}

By an accurate analysis of (\ref{2nd}), one can see that it is not
general covariant
in a curved space-time,
because of ${}^*\!F$ is not a scalar $2$-form, but it is a relative scalar
$2$-form with weight $-1$, due to the anti-symmetrical symbol.
Now we assume that the manifold is endowed with a
connection to use it in order to cast (\ref{2nd}) in a covariant way. This is
done by substituting the exterior derivative by the covariant one
\begin{equation}
d{}^*\!F \rightarrow {\cal D}{}^*\!F = \frac{1}{3!}\left(
\partial_\alpha {}^*\! F_{\beta\gamma} +
\Gamma^\rho_{\rho\alpha}{}^*\! F_{\beta\gamma}
\right)\delta^{\alpha\beta\gamma}_{\mu\nu\omega} \,
dx^\mu \wedge dx^\nu \wedge dx^\omega,
\label{covex}
\end{equation}
where $\delta^{\alpha\beta\gamma}_{\mu\nu\omega}$ is the generalized Kronecker
delta.
The covariant exterior derivative in (\ref{covex})  takes into account
that
${}^*\!F_{\alpha\beta} =
\frac{1}{2}\varepsilon_{\alpha\beta\gamma\delta}F^{\gamma\delta}$
is a relative $(0,2)$ tensor with weight $-1$.
One can check that ${\cal D}{}^*\!F$ is a relative scalar $3$-form with
weight $-1$.
We have then the following
covariant generalization of (\ref{2nd})
\begin{equation}
{\cal D} {}^*\!F = 4\pi {}^*\! J.
\label{2ndcov}
\end{equation}
Equations (\ref{1st}) are already in a general
covariant form in any differentiable
manifold.

Taking the covariant exterior derivative in both sides of (\ref{2ndcov}) we
get
\begin{equation}
4\pi {\cal D} {}^*\! J = \frac{1}{4!}\left(
\partial_\lambda\Gamma^\rho_{\rho\mu} \right) {}^*\! F_{\nu\omega}
\delta^{\lambda\mu\nu\omega}_{\alpha\beta\gamma\delta}\,
dx^\alpha \wedge dx^\beta \wedge dx^\gamma \wedge dx^\delta,
\label{conserv}
\end{equation}
and to have a generalized conservation condition for the current we need that
\begin{equation}
\partial_\lambda \Gamma^\rho_{\rho\mu} -
\partial_\mu \Gamma^\rho_{\rho\lambda} = 0,
\end{equation}
which has locally  as general solution:
\begin{equation}
\Gamma^\rho_{\rho\mu} = \partial_\mu f(x).
\label{condit}
\end{equation}
Using that $\left\{^\rho_{\rho\mu} \right\}=\partial_\mu \ln \sqrt{-g}$,
equation (\ref{condit})
will have general solution only if the trace of the torsion
tensor is derivable from a scalar potential, or, that is equivalent,
that the space-time admits a compatible volume-form (Proposition 1).
 In this case
$f(x)=\ln \left(e^{2\Theta}\sqrt{-g}\right)$.

One can ask now if it is possible to obtain the non-homogeneous equations
(\ref{2ndcov}) from an action principle. We know that in Minkowski
space-time,
the non-homogeneous equations are gotten from the following action
\begin{equation}
S = -\int \left(4\pi{}^*\!J\wedge A +\frac{1}{2} F \wedge {}^*\!F\right).
\label{actmink}
\end{equation}
The action (\ref{actmink}) can be cast in a covariant way by using
the modified Hodge $({}^*)$ operator (Proposition 3). In this case one gets the
following coordinate expression for the generally covariant generalization of
(\ref{actmink})
\begin{equation}
S = \int d^4x \, e^{2\Theta} \sqrt{-g}\left(
-\frac{1}{4}F_{\alpha\beta} F^{\alpha\beta} + 4\pi J^\alpha A_\alpha
\right).
\label{actu4}
\end{equation}
It is easy to check that we can obtain equations (\ref{2ndcov}) from the
action (\ref{actu4}). Under (\ref{condit}),
the covariant equation (\ref{2ndcov}) can be obtained directly from
(\ref{2nd}) by using the modified Hodge operator.
We can check also that equations (\ref{1stc}) and
(\ref{2ndcov}) are invariant under the usual $U(1)$ gauge transformation
\begin{equation}
A_\mu \rightarrow A_\mu + \partial_\mu \varphi.
\label{u1gauge}
\end{equation}
We would like to stress the importance of the generalized conservation
condition  to guarantee the gauge invariance of the
action (\ref{actu4}).

The use of the compatible volume element have brought two main modifications
to the problem of Maxwell fields on Riemann-Cartan space-times. First,
it is clear from (\ref{actu4}) that gauge fields can interact with torsion
without destroying gauge invariance, and second, there is no difference
if one starts from the action formulation or from the equations of
motion.  It is
important to stress also that the necessary condition  to a
Riemann-Cartan manifold admit a compatible volume element arises as an
``integrability'' condition for Maxwell's
 equations on Riemann-Cartan manifolds.

\section{Final Remarks}

The consequences of using compatible volume elements for minimal action
principles for field theory is now under investigation. For the
Einstein-Cartan theory, it is already known that the new volume element
will modify the dynamical equations for the space-time geometry.
Propagation of torsion will
be possible, and not only fermion fields
will interact with the non-riemannian structure of space-time. The theory
become more consistent, since we get the same results starting
from the action formulation or from the equations of motion. It is very
interesting that the necessary condition of Proposition 2 arises as
integrability condition for the matter field equations  in Riemann-Cartan
space-times[\citen{saa1,saa2}].
 The compatible
volume element may also provide a geometrical interpretation for the
dilaton gravity, which comes from string theory.

\section*{Acknowledgements}

The author is grateful to  Josif Frenkel and Jos\'e Carlos
Brunelli. This
work was supported by Funda\c c\~ao de Amparo \`a Pesquisa do Estado de
S\~ao Paulo.


\newpage

\begin{thebibliography}{9}

\bibitem{hehl} F.W. Hehl, P. von der Heyde, G.D. Kerlick, and J.M. Nester,
{\em General Relativity with Spin and Torsion: Foundations and Prospects},
Rev. Mod. Phys. {\bf 48}  (1976) 393.

\bibitem{kob} S. Kobayashi and K. Nomizu, {\em Foundations of Differential
Geometry}, vol. 1, John Willey, New York, 1963.

\bibitem{nakahara} M. Nakahara, {\em Geometry, Topology and Physics},
Adam Hilger, Bristol, 1990.

\bibitem{saa1} A. Saa, {\em On Minimal Coupling in Riemann-Cartan Space-times},
to be published in Mod.Phys.Lett. A.

\bibitem{saa2}A. Saa, {\em Gauge Fields on Riemann-Cartan Space-times}, to
be published in Int.J.Mod.Phys. A.

\bibitem{callan}C.G. Callan, D. Friedan, E.J. Martinec, and M.J. Perry,
{\em Strings in Background Fields}, Nucl.Phys. {\bf B262} (1985) 593.

\bibitem{green}M.B. Green, J.H. Schwarz, and E. Witten,
{\em Superstring Theory}, sect. 3.4, Cambridge University Press, 1987.

\bibitem{saa3}A. Saa, {\em Strings in Background Fields and Einstein-Cartan
theory of Gravity}, Pre-print IFUSP/P-1066 (HEP-TH/9307095), to be
published.

\end{thebibliography}
\end{document}